\documentstyle[11pt]{article}
\begin{document}
\baselineskip 19pt

\title{The Perturbative Quantization of Gravity}

\author{
A.Y. Shiekh\thanks{\tt shiekh@ictp.trieste.it}
\\
International Centre for Theoretical Physics,
Miramare, Trieste, Italy
}

\date{28th June 1994}

\maketitle

\begin{abstract}
A suggestion is made for quantizing gravity
perturbatively, and is illustrated for the example of a
massive scalar field with gravity.
\end{abstract}

\section{\bf The Scheme}

        This short talk is about the renormalizability of
Einstein gravity\footnote{This talk was given at the
Henri Poincar{\'e} seminar held at Protvino Russia,
and is
a condensed report of the work ``Orthodox Gravity'',
presently under submission}. It is {\it NOT} renormalizable so this might
indeed be a very short talk!

        However the work might best be viewed as a look back over
failed campaigns of the past. We are forever rushing ahead
with new ideas, that perhaps sometimes we lose
perspective. So let us review; and if we happen to stumble
upon a trail missed before, all the better.

        Renormalization is all about reabsorbing infinities into
the starting Lagrangian [Ramond, 1990; Collins, 1989].
This can fail in one of two ways:

\vskip .5cm

{\it
I) The terms to be reabsorbed are not present in the
original Lagrangian.

\vskip .5cm

II) One might keep extending the starting Lagrangian to
take up the counter terms and end up with an infinite
number of terms and associated undetermined coupling
constants. The theory is then technically renormalized,
but has lost all predictive content.
}

\vskip .5cm

        This short review complete, we can look at the problems
of quantizing Einstein gravity with a massive scalar field
[Veltman, 1976].

        Starting from the example of a free scalar field with
gravity described by the classical Lagrangian in Euclidean
space:

$$L=-\sqrt g\left( {R+{\textstyle{1 \over 2}}g^{\mu \nu
}\partial _\mu \phi \partial _\nu \phi +{\textstyle{1
\over 2}}m^2\phi ^2} \right)$$

\rightline{\it (using units where $16\pi \,G=1$, $c=1$) }

\noindent one proceeds by extracting the Feynman rules (of
which there are an infinite number, although only a finite
number are used to any finite loop order). At one loop one
already encounters infinities that cannot be reabsorbed
into the starting Lagrangian. Einstein gravity is not
renormalizable.

        This very malady suggests its own solution: namely to
extend the original Lagrangian so that symmetry will
ensure that the counter terms all fall within the original
Lagrangian (the second line carries all the higher
derivatives):

$$
\begin{array}{lll}
L_0=-\sqrt {g_0}
\\
\left( \matrix{-2\Lambda
_0+R_0+{\textstyle{1 \over 2}}p_0^2+{\textstyle{1 \over
2}}m_0^2\phi _0^2+{\textstyle{1 \over {4!}}}\phi
_0^4\lambda _0(\phi _0^2)+p_0^2\phi _0^2\kappa (\phi
_0^2)+R_0\phi _0^2\gamma _0(\phi _0^2)\cr
  +p_0^4a_0(p_0^2,\phi _0^2)+R_0p_0^2b_0(p_0^2,\phi
_0^2)+R_0^2c_0(p_0^2,\phi _0^2)+R_{0\mu \nu }R_0^{\mu \nu
}d_0(p_0^2,\phi _0^2)+...\cr} \right)
\end{array}
$$

\rightline{\it (where $p_0^2$ is shorthand for
 $g_o^{\mu \nu }\partial _\mu \phi _0
\partial _\nu \phi _0$) }

\noindent which then renormalizes to:

$$
\begin{array}{lll}
L=-\sqrt g
\\
\left( \matrix{-2\Lambda +R+{\textstyle{1
\over 2}}p^2+{\textstyle{1 \over 2}}m^2\phi
^2+{\textstyle{1 \over {4!}}}\phi ^4\lambda (\phi
^2)+p^2\phi ^2\kappa (\phi ^2)+R\phi ^2\gamma (\phi ^2)\cr
  +p^4a(p^2,\phi ^2)+Rp^2b(p^2,\phi ^2)+R^2c(p^2,\phi
^2)+R_{\mu \nu }R^{\mu \nu }d(p^2,\phi ^2)+...\cr}
\right)
\end{array}
$$

        Now all is lost from the start, for there are an infinite
number of arbitrary coupling constants. We have played a
futile game, but at least we can `formally' renormalize
this particular gravity theory. No progress can be made
with this approach unless there is more physics at hand.

        It is at this point that we make recourse to `desperate
measures'. As a first strike one might simply cut back to
the desired end theory (supposedly Einstein gravity) by
setting undesired renormalized coupling constants to zero.
But perhaps we can do better, and argue our way forward.

        Paradoxically perhaps for science, some lines of
reasoning blossom and wane with the fashion of the day.
Einstein himself (after several attempts) proposed $R$
 gravity, abandoning higher derivative gravity on the
grounds that such theories in general violate causality.
As a matter of necessity we will follow the same route,
despite the modern acceptance of these pathologies (as in
string theory). We go on to also abandon $R$
 coupling terms on the grounds that they violate the
equivalence principle, and further insist that in the flat
space-time limit the resultant theory is renormalizable in
the `traditional sense', so abandoning such terms as $\phi
^6$.

        These criteria guide us to set the renormalized coupling
of all but a finite number of terms to zero, leading to:

$$L=-\sqrt g\left( {-2\Lambda +R+{\textstyle{1 \over
2}}p^2+{\textstyle{1 \over 2}}m^2\phi ^2+{\textstyle{1
\over {4!}}}\lambda \phi ^4} \right)$$

        But even if we get to this stage, we have worries about
the renormalization group pulling the coupling constants
around. This is an open point to which I feel one of three
things might happen:

\vskip .5cm

{\it
I) The couplings, set to zero at a low energy scale,
might reappear around the Plank scale. Whether the
resulting theory then makes sense at this scale is a
matter for dispute.

\vskip .5cm

II) Certain extra coupling constants should be zero,
in-order that the beta functions be zero, ensuring that
all the couplings set to zero stay zero. This consistency
condition could be the basis of a unification scheme.

\vskip .5cm

III) Some work [Culumovic et al., 1990; Leblanc et al.,
1991] suggests that for traditionally unrenormalizable
theories a consistency condition arises that fixes the
renormalization group parameter, supposedly at the Plank
scale for gravity. This idea is very tentative and not
conclusive.
}

\vskip .5cm

        In some sense we have come to the end of the trail and
the scheme for quantizing gravity is now presented.
However, I would like to continue on a diverse, but
related track.

\vskip 1cm

\section{\bf The Method}

        There is an `unsung hero' in the guise of operator
regularization [McKeon and Sherry, 1987; McKeon et al.,
1987; McKeon et al., 1988; Mann, 1988; Mann et al., 1989;
Culumovic et al., 1990; Shiekh, 1990]. Operator
regularization is an n-loop generalization of the one loop
Zeta function analytic continuation technique [Salam and
Strathdee, 1975; Dowker and Critchley, 1976; Hawking,
1975]. As such, quantum field theory is finite from start
to finish under this technique. Since one does not change
the number of space-time dimensions, it even maintains
more symmetries than dimensional regularization and
calculations are almost identical (c.f. the added
complexity of the Pauli-Villars regulator).

        I am a keen supporter of the operator regularization
technique, although I will level a criticism at its
present formulation, which consists of the replacement of
a divergent part by the analytic continuation given by:

$$\Omega ^{-m}=\mathop {\lim}\limits_{\varepsilon \to
0}{{d^n} \over {d\varepsilon ^n}}\left( {{{\varepsilon ^n}
\over {n!}}\Omega ^{-\varepsilon -m}} \right)$$

\noindent where $n$ is chosen sufficiently large to eliminate the infinities
(the loop order is sufficient). Its use will be explicitly
illustrated later. Actually, operator regularization is a
bit of a misnomer, since it need not be applied to an
operator and does not just regulate, but also renormalizes
all in one.

        However, under this form of the method all theories are
finite and predictive (gravity included). A little playing
shows that the above is simply an automated system for
minimal subtraction, and this realized, the general form
is easily located, and is given by:

$$\Omega ^{-m}=\mathop {\lim}\limits_{\varepsilon \to
0}{{d^n} \over {d\varepsilon ^n}}\left( {\left( {1+\alpha
_1\varepsilon +...+\alpha _n\varepsilon ^n}
\right){{\varepsilon ^n} \over {n!}}\Omega ^{-\varepsilon
-m}} \right)$$

\rightline{\it (the alphas being ambiguous)}

        This form is not too powerful, and gravity must again be
dealt with the desperate measures of before.

        With all this machinery in hand one might want to walk
through a simple example of a divergent one loop diagram.
So begin with an investigation of a massive scalar theory
in its own induced gravitational field, described by the
Lagrangian in Euclidean space given by:

$$L=-\sqrt g\;\left( {-2\Lambda +R+{\textstyle{1 \over
2}}g^{\mu \nu }\partial _\mu \phi \partial _\nu \phi
+{\textstyle{1 \over 2}}m^2\phi ^2+{\textstyle{1 \over
{4!}}}\lambda \phi ^4} \right)$$

        The Euclidean Feynman rules (of which there are an
infinite number) we explicitly list; the gauged graviton
propagator being derived from the gravitational, $R$,
Lagrangian [Veltman, 1976], in the harmonic
gauge\footnote{Postscript figures of the Feynman rules and diagrams
are available from the author at {\tt shiekh@ictp.trieste.it}}:

$$FIG.1={{\delta _{\mu \alpha }\delta _{\nu \beta }+\delta _{\mu
\beta }\delta _{\nu \alpha }-\delta _{\mu \nu }\delta
_{\alpha \beta }} \over {p^2}}$$

\noindent The scalar propagator is given by:

$$FIG.2 ={1 \over {p^2+m^2}}$$

\noindent and the first interaction vertex by:

$$FIG.3={\textstyle{1 \over 2}}\delta _{\mu \nu }\left( {p\cdot
q-m^2} \right)-p_\mu q_\nu $$

\noindent etc., using units where $\hbar =1$.

        Although there are an infinite number of Feynman
diagrams, only a finite number are used to any finite loop
order.

\vskip .5cm

\noindent {\bf Divergent one loop diagram example:}

        Set about a one loop investigation with matter particles
on the external legs:

$$
\begin{array}{ccc}
FIG.4=
\\
\int_{-\infty }^\infty  {{{d^4l} \over
{\left( {2\pi } \right)^4}}}\left( {{1 \over {l^2+m^2}}}
\right)\left( {{{\delta _{\mu \alpha }\delta _{\nu \beta
}+\delta _{\mu \beta }\delta _{\nu \alpha }-\delta _{\mu
\nu }\delta _{\alpha \beta }} \over {\left( {l+p}
\right)^2}}} \right)
\\
  \left( {{\textstyle{1 \over 2}}\delta _{\mu \nu }\left(
{p\cdot l-m^2} \right)-p_\mu l_\nu } \right)\left(
{{\textstyle{1 \over 2}}\delta _{\alpha \beta }\left(
{p\cdot l-m^2} \right)-p_\alpha l_\beta } \right)
\end{array}
$$

\noindent Expand out the indices to yield:

$$=\int_{-\infty }^\infty  {{{d^4l} \over {\left( {2\pi }
\right)^4}}}{1 \over {l^2+m^2}}{1 \over {\left( {l+p}
\right)^2}}\left( {p^2l^2+2m^2p\cdot l-2m^4} \right)$$

\noindent and then introduce the standard Feynman parameter
 `trick':
$$
\begin{array}{lll}
{1 \over {D_1^{a_1}D_2^{a_2}...D_k^{a_k}}}=
{{\Gamma
(a_1+a_2+...a_k)} \over {\Gamma (a_1)\Gamma (a_2)...\Gamma
(a_k)}}\int_0^1 {...}\int_0^1 {dx_1...dx_k}{{\delta
(1-x_1-...x_k)x_1^{a_1-1}...x_k^{a_k-1}} \over {\left(
{D_1x_1+...D_kx_k} \right)^{a_1+...a_k}}}
\end{array}
$$

\noindent to yield:

$$=\int_{-\infty }^\infty  {{{d^4l} \over {\left( {2\pi }
\right)^4}}\int_0^1 {dx}}{{p^2l^2+2m^2p\cdot l-2m^4} \over
{\left[ {l^2+m^2x+p^2\left( {1-x} \right)+2l\cdot p\left(
{1-x} \right)} \right]^2}}$$

\noindent Remove divergences using operator regularization:

$$\Omega ^{-m}=\mathop {\lim}\limits_{\varepsilon \to
0}{{d^n} \over {d\varepsilon ^n}}\left( {\left( {1+\alpha
_1\varepsilon +...+\alpha _n\varepsilon ^n}
\right){{\varepsilon ^n} \over {n!}}\Omega ^{-\varepsilon
-m}} \right)$$

\noindent $n$ being chosen sufficiently large to cancel the infinities.
For the case in hand $n=1$ is adequate.

$$\Omega ^{-2}=\mathop {\lim}\limits_{\varepsilon \to
0}{d \over {d\varepsilon }}\left( {\left( {1+\alpha
\varepsilon } \right)\varepsilon \Omega ^{-\varepsilon
-2}} \right)$$

\noindent This leads to:

$$=\int_0^1 {dx}\;\mathop {\lim}\limits_{\varepsilon \to
0}{d \over {d\varepsilon }}\int_{-\infty }^\infty
{{{d^4l} \over {\left( {2\pi } \right)^4}}}\left(
{\varepsilon \left( {1+\alpha \varepsilon }
\right){{p^2l^2+2m^2p\cdot l-2m^4} \over {\left[
{l^2+m^2x+p^2\left( {1-x} \right)+2l\cdot p\left( {1-x}
\right)} \right]^{\varepsilon +2}}}} \right)$$

\noindent Then performing the momentum integrations using [Ramond,
1990]:

$$
\begin{array}{lll}
\int_{-\infty }^\infty  {{{d^{2\omega }l} \over {\left(
{2\pi } \right)^{2\omega }}}}{1 \over {\left(
{l^2+M^2+2l\cdot p} \right)^A}}={1 \over {(4\pi )^\omega
\Gamma (A)}}{{\Gamma (A-\omega )} \over
{(M^2-p^2)^{A-\omega }}}
\end{array}
$$

$$
\begin{array}{lll}
\int_{-\infty }^\infty  {{{d^{2\omega }l} \over {\left(
{2\pi } \right)^{2\omega }}}}{{l_\mu } \over {\left(
{l^2+M^2+2l\cdot p} \right)^A}}=-{1 \over {(4\pi )^\omega
\Gamma (A)}}p_\mu {{\Gamma (A-\omega )} \over
{(M^2-p^2)^{A-\omega }}}
\end{array}
$$

$$
\begin{array}{lll}
\int_{-\infty }^\infty  {{{d^{2\omega }l} \over
{\left( {2\pi } \right)^{2\omega }}}}{{l_\mu l_\nu } \over
{\left( {l^2+M^2+2l\cdot p} \right)^A}}=
  {1 \over {(4\pi )^\omega \Gamma (A)}}\left[ {p_\mu
p_\nu {{\Gamma (A-\omega )} \over {(M^2-p^2)^{A-\omega
}}}+{{\delta _{\mu \nu }} \over 2}{{\Gamma (A-\omega -1)}
\over {(M^2-p^2)^{A-\omega -1}}}} \right]
\end{array}
$$

\noindent yields the finite expression:

$$={1 \over {\left( {4\pi } \right)^2}}\int_0^1
{dx}\;\mathop {\lim}\limits_{\varepsilon \to 0}{d \over
{d\varepsilon }}\left( {{{\varepsilon \left( {1+\alpha
\varepsilon } \right)} \over {\Gamma (\varepsilon
+2)}}\left( \matrix{{{p^4(1-x)^2\Gamma (\varepsilon )}
\over {\left[ {m^2x+p^2x\left( {1-x} \right)}
\right]^\varepsilon }}+2{{p^2\Gamma (\varepsilon -1)}
\over {\left[ {m^2x+p^2x\left( {1-x} \right)}
\right]^{\varepsilon -1}}}\cr
  -2{{m^2p^2(1-x)\Gamma (\varepsilon )} \over {\left[
{m^2x+p^2x\left( {1-x} \right)} \right]^\varepsilon
}}-2{{m^4\Gamma (\varepsilon )} \over {\left[
{m^2x+p^2x\left( {1-x} \right)} \right]^\varepsilon }}\cr}
\right)} \right)$$

\noindent Doing the $\varepsilon $ differential using:

$$\mathop {\lim}\limits_{\varepsilon \to 0}{d \over
{d\varepsilon }}\left( {{{\varepsilon \left( {1+\alpha
\varepsilon } \right)} \over {\Gamma (\varepsilon
+2)}}\left( {a{{\Gamma (\varepsilon )} \over {\chi
^\varepsilon }}+b{{\Gamma (\varepsilon -1)} \over {\chi
^{\varepsilon -1}}}} \right)} \right)=-a+\left( {a-b\chi }
\right)\left( {\alpha -\ln (\chi )} \right)$$

\noindent yields:

$$={1 \over {\left( {4\pi } \right)^2}}\int_0^1 {dx}\left(
\matrix{\left( {\left( {2m^4+2m^2p^2-p^4}
\right)+p^4x\left( {4-3x} \right)} \right)\left( {\ln
\left( {m^2x+p^2x(1-x)} \right)-\alpha } \right)\cr
  +2m^4+2m^2p^2-p^4-p^2x\left( {2m^2-2p^2+p^2x}
\right)\cr} \right)$$

\noindent and finally performing the $x$
 integration gives rise to the final result in Euclidean
space, namely:

$$FIG.4={{m^4} \over {\left( {4\pi } \right)^2}}\left(
\matrix{\left( {3+2{{p^2} \over {m^2}}+{{m^2} \over
{p^2}}} \right)\ln (1+{{p^2} \mathord{\left/ {\vphantom
{{p^2} {m^2}}} \right. \kern-\nulldelimiterspace}
{m^2}})-1-{5 \over 2}{{p^2} \over {m^2}}\cr
  -{1 \over 6}{{p^4} \over {m^4}}+2\left( {1+{{p^2} \over
{m^2}}} \right)\left( {\ln ({{m^2} \mathord{\left/
{\vphantom {{m^2} {\mu ^2}}} \right.
\kern-\nulldelimiterspace} {\mu ^2}})-\alpha } \right)\cr}
\right)$$

\noindent where there is no actual divergence at $p=0$, and it
should be commented that the use of a computer mathematics
package can in general greatly reduced `calculator'
fatigue. The renormalization group factor $\mu $
 appears on dimensional grounds.

\section {\bf Comments}

        The mere existence of a finite perturbative formulation
of quantum gravity might give us reason to return once
again to Einstein gravity in a quantum context. It would
at least be a start.

\section {\bf Acknowledgements}

        I should like to thank the organizers of the Protvino
seminar for a friendliness never to be forgotten; members
of a country in the flux of change.

\section {\bf References}

{\small
\noindent 1)    P. Ramond, {\it ``Field Theory: A Modern Primer''},
2nd Ed (Addison-Wesley, 1990).

\noindent 2)    J. Collins,
{\it ``Renormalization''}, Cambridge University Press, London,
1984.

\vskip .5cm

\noindent 3)    M.J.G. Veltman, {\it `Quantum Theory of Gravitation'}, Les
Houches XXVIII, ``Methods In Field Theory', pp. 265-327,
eds. R. Ballan and J. Zinn-Justin, (North-Holland,
Amsterdam, 1976).

\noindent 4)    L. Culumovic, D.G.C. McKeon and T.N. Sherry, {\it `Operator
regularization and massive Yang-Mills theory'}, Can. J.
Phys., 68, 1990, 1149-.

\noindent 5)    M. Leblanc, R.B. Mann, D.G.C.
McKeon and T.N. Sherry, {\it `The Finite Effective Action in
the Non-Linear Sigma Model with Torsion to Two-Loop
Order'}, Nucl. Phys. B349, 1991, 494-.

\vskip .5cm

\noindent 6)    A. Salam and J. Strathdee, {\it `Transition Electromagnetic
Fields in Particle Physics'}, Nucl. Phys. B90, 1975, 203-.

\noindent 7)    J. Dowker and R. Critchley, {\it `Effective Lagrangian and
energy-momentum tensor in de sitter space'}, Phys. Rev.,
D13, 1976, 3224-.

\noindent 8)    S. Hawking, {\it `Zeta Function
Regularization of Path Integrals in Curved Space'}, Commun.
Math. Phys., 55, 1977, 133-.

\vskip .5cm

\noindent 9)    D. McKeon and T. Sherry, {\it `Operator Regularization of
Green's Functions'}, Phys. Rev. Lett., {\bf 59}, 1987, 532-.

\noindent 10) D. McKeon and T. Sherry, {\it `Operator Regularization and
one-loop Green's functions'}, Phys. Rev., {\bf D35}, 1987, 3854-.

\noindent 11)   D. McKeon, S. Rajpoot and T. Sherry, {\it `Operator
Regularization with Superfields'}, Phys. Rev., {\bf D35}, 1987,
3873-.

\noindent 12)   D. McKeon, S. Samant and T. Sherry, {\it `Operator
regularization beyond lowest order'}, Can. J. Phys., {\bf 66},
1988, 268-.

\noindent 13)   R. Mann., {\it `Zeta function regularization of
Quantum Gravity'}, In Proceedings of the cap-nserc Summer
Workshop on Field Theory and Critical Phenomena. Edited by
G. Kunstatter, H. Lee, F. Khanna and H. Limezawa, World
Scientific Pub. Co. Ltd., Singapore, 1988, p. 17-.

\noindent 14)   R.
Mann, D. McKeon, T. Steele and T. Tarasov, {\it `Operator
Regularization and Quantum Gravity'}, Nucl. Phys., {\bf B311},
1989, 630-.

\noindent 15)   L. Culumovic, M. Leblanc, R. Mann, D.
McKeon and T. Sherry, {\it `Operator regularization and
multiloop Green's functions'}, Phys. Rev., {\bf D41}, 1990, 514-.

\noindent 16)   A. Shiekh, {\it `Zeta-function regularization of
quantum field theory'}, Can. J.Phys., {\bf 68}, 1990, 620-.

\null
}

\end{document}